\RequirePackage{fix-cm}
\documentclass[smallextended]{svjour3}       % onecolumn (second format)
\smartqed  % flush right qed marks, e.g. at end of proof
\usepackage{graphicx}
\usepackage{amsmath}
\usepackage{amsfonts}
\usepackage{tikz}
\usepackage[draft]{changes}

\newcommand{\IversonBracket}[1]{[\negthinspace[{#1}]\negthinspace]}

\newcommand{\RR}{{\mathbb R}}
\newcommand{\bx}{\mathbf{x}}
\newcommand{\by}{\mathbf{y}}
\newcommand{\bc}{{\bf c}}
\newcommand{\bn}{{\bf 0}}
\newcommand{\bs}{{\bf s}}
\newcommand{\bw}{{\bf w}}
\newcommand{\bz}{{\bf z}}
\newcommand{\bb}{{\bf b}}
\newcommand{\bena}{\mathbf{1}}
\newcommand{\benat}{\bena\kern-0.1em^{\raisebox{-1pt}{$\scriptstyle\top$}}}
\newcommand{\eps}{\varepsilon}

\begin{document}

\title{Concavity of cardinality-constrained interval stabbing and interval cover problems
\thanks{This study was supported by the research programmes P2-0103 and P2-0209 
of the Slovenian Research and Innovation Agency}
}

\titlerunning{Concavity of cardinality-constrained interval problems}        % if too long for running head

\author{Janez Brank         \and
        Tomaž Hočevar %etc.
}


\institute{Janez Brank \at
              Jožef Stefan Institute, Ljubljana, Slovenia \\
              \email{janez.brank@ijs.si}           
           \and
           Tomaž Hočevar \at
              University of Ljubljana, Faculty of Computer and Information Science, Ljubljana, Slovenia\\
              \email{tomaz.hocevar@fri.uni-lj.si}           
}

\date{Received: date / Accepted: date}

%%%%% Suppress journal specifics
\def\makeheadbox{}  % Remove journal heading / manuscript box
\date{}  % Suppress Received / Accepted line
\pagestyle{plain}  % Simple page numbers, no running journal headers

\maketitle

\begin{abstract}

We consider the cardinality-constrained interval stabbing problem, in which the goal is to maximize the total weight of intervals hit by at least one of at most $k$ selected points. Although this problem does not admit a simple greedy solution, we show that its optimal value is a concave function of $k$. We establish this result using a linear programming formulation whose relaxation always admits an optimal solution with integer values. This allows us to prove concavity for the original problem. We further extend the result to the related interval cover problem. These concavity properties provide a basis for applying Lagrangian optimization using algorithms for the corresponding unconstrained problems.

\end{abstract}

\section{Introduction}

Interval stabbing is a classic problem in combinatorial optimization. Given a set of $N$ intervals
$[\ell_j, r_j]$ we need to select a minimum-cardinality set of points $\bx$ (stabbing locations) such that every interval contains at least one selected point. In this paper, we consider a cardinality-constrained variant, where we are allowed to select at most $k$ stabbing points. The intervals are associated with positive weights $w_j$ and we aim to maximize the total weight of intervals that are stabbed at least once: $f(k)$.

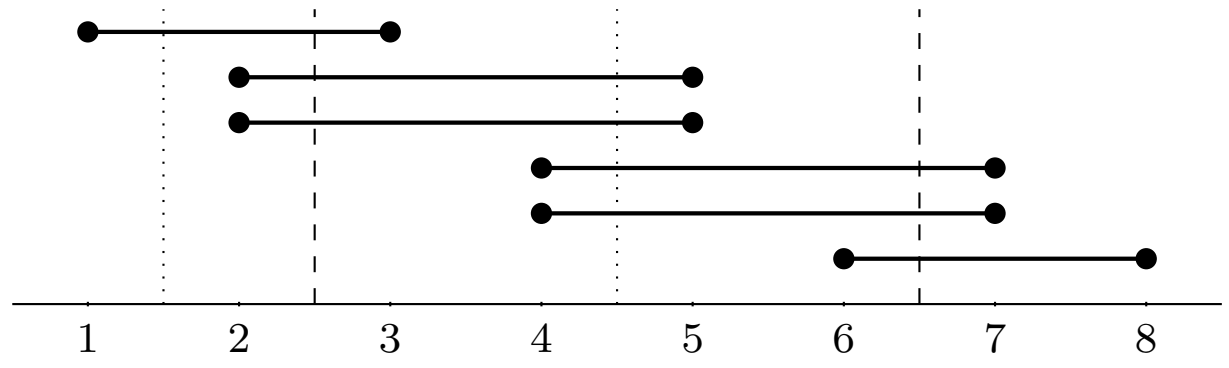
\begin{figure}
\centering
\begin{tikzpicture}[x=1cm, y=0.3cm]
    % Intervals
    \foreach \a/\b/\y/\lab in {
        1/3/6.5/I_1,
        2/5/5.5/I_2,
        2/5/4.5/I_3,
        4/7/3.5/I_4,
        4/7/2.5/I_5,
        6/8/1.5/I_6
    } {
        \draw[thick] (\a,\y) -- (\b,\y);
        \fill (\a,\y) circle (2pt);
        \fill (\b,\y) circle (2pt);
        %\node[left] at (0.8,\y) {$\lab$};
    }

    % Stabbing points
    \draw[dashed] (2.5,0.5) -- (2.5,7.0);
    \draw[dashed] (6.5,0.5) -- (6.5,7.0);
    
    % Greedy choice
    \draw[dotted] (4.5,0.5) -- (4.5,7.0);
    \draw[dotted] (1.5,0.5) -- (1.5,7.0);

    % Horizontal axis
    \draw[-] (0.5,0.5) -- (8.5,0.5);

    % Axis labels
    \foreach \x in {1,...,8} {
        \draw (\x,0.45) -- (\x,0.55);
        \node[below] at (\x,0.4) {$\x$};
    }
    
    % Stabbing labels
    %\node[red, above] at (2.5,5.5) {$2.5$};
    %\node[red, above] at (6.5,5.5) {$6.5$};
\end{tikzpicture}

\caption{
An unweighted counter-example to a greedy algorithm for cardinality-constrained interval stabbing with $k=2$.
The greedy strategy (dotted lines) would make the first stab in the middle to stab four intervals. However, the two remaining intervals are disjoint, so a second point cannot stab both. An optimal solution (dashed lines) can stab all six intervals (three with each stab).
}
\label{fig:greedy}
\end{figure}

The classical problem of stabbing all intervals with the minimum number of stabs admits a simple greedy algorithm: repeatedly select the leftmost endpoint among the intervals and remove all stabbed intervals \cite{erickson2019algorithms}.
However, a greedy strategy does not directly extend to the cardinality-constrained problem. Selecting a point that would stab a set of intervals with the largest sum of weights does not always lead to an optimal solution (see Figure~\ref{fig:greedy} for a counter-example). To solve the problem optimally, we would use a dynamic programming approach. 

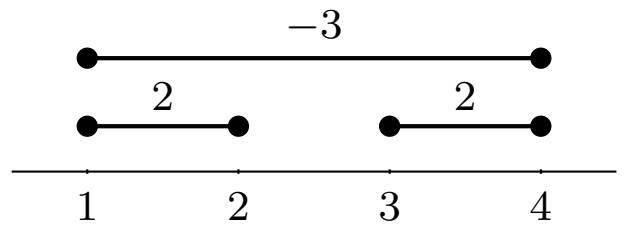
\begin{figure}[hb]
\centering
\begin{tikzpicture}[x=1cm, y=0.3cm]
    % Intervals
    \foreach \a/\b/\y/\w in {
        1/2/1.5/2,
        3/4/1.5/2,
        1/4/3.0/-3
    } {
        \draw[thick] (\a,\y) -- (\b,\y);
        \fill (\a,\y) circle (2pt);
        \fill (\b,\y) circle (2pt);
        \node[above] at ({(\a+\b)/2},\y) {$\w$};
    }

    % Horizontal axis
    \draw[-] (0.5,0.5) -- (4.5,0.5);

    % Axis labels
    \foreach \x in {1,...,4} {
        \draw (\x,0.45) -- (\x,0.55);
        \node[below] at (\x,0.4) {$\x$};
    }
\end{tikzpicture}

\caption{
An example with negative weights. The optimal values $f(k)$, for $k=0,\ldots,3$, are $\{0,0,1,1\}$, which are non-decreasing but not concave. With only one selected stabbing point, it is optimal to avoid stabbing any interval, whereas two points allow both positive-weight intervals to be stabbed, overcoming the negative contribution of the longest interval.
}
\label{fig:negative}
\end{figure}

The case of negative weights $w_j$ is fundamentally different. Although the optimal value function $f(k)$ remains non-decreasing, since additional points can be selected without stabbing any interval, it might not be concave, as illustrated in Figure~\ref{fig:negative}. Throughout this paper, we therefore restrict our attention to strictly positive weights.

Let $f(k)$ denote the maximum total weight of intervals that can be stabbed using at most $k$ points. Clearly, $f(k)$ is non-decreasing, since allowing additional stabbing points cannot decrease the optimal value. More surprisingly, we show that $f(k)$ is also concave, which means that the marginal benefit of adding a stabbing point is non-increasing. This behavior is often associated with problems that admit greedy solutions, where components of the solution are selected in order of decreasing marginal benefit, which is not the case here. It is therefore notable that concavity holds even though the cardinality-constrained interval stabbing problem does not admit a simple greedy algorithm.

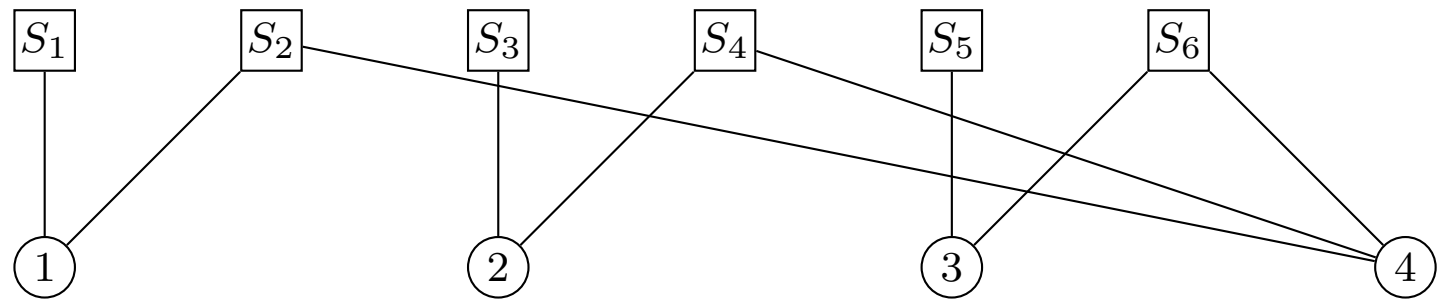
\begin{figure}
\centering
\begin{tikzpicture}[
  set/.style={draw, rectangle, minimum size=4mm, inner sep=1pt},
  pt/.style={draw, circle, minimum size=4mm, inner sep=1pt},
  every edge/.style={draw, thick}
]
  % sets (top row), all of weight 1
  \node[set] (s1) at (0,1.5)  {$S_1$};
  \node[set] (s2) at (1.5,1.5) {$S_2$};
  \node[set] (s3) at (3,1.5)  {$S_3$};
  \node[set] (s4) at (4.5,1.5) {$S_4$};
  \node[set] (s5) at (6,1.5)  {$S_5$};
  \node[set] (s6) at (7.5,1.5) {$S_6$};

  % points (bottom row)
  \node[pt] (u0) at (0,0) {$1$};
  \node[pt] (u1) at (3,0) {$2$};
  \node[pt] (u2) at (6,0) {$3$};
  \node[pt] (u3) at (9,0)  {$4$};

  % S_1={0}, S_2={0,3}, S_3={1}, S_4={1,3}, S_5={2}, S_6={2,3}
  \draw (s1) -- (u0);
  \draw (s2) -- (u0);  \draw (s2) -- (u3);
  \draw (s3) -- (u1);
  \draw (s4) -- (u1);  \draw (s4) -- (u3);
  \draw (s5) -- (u2);
  \draw (s6) -- (u2);  \draw (s6) -- (u3);
\end{tikzpicture}

\caption{
An unweighted counter-example of concavity of the cardinality-constrained hitting set problem with a set of $n=4$ points $U=\{1,2,3,4\}$ and a collection of $N=6$ sets $S = \{\{1\}, \{1,4\}, \{2\}, \{2,4\}, \{3\}, \{3,4\} \}$.
The optimal solutions $f(k)$ for $k=0,\ldots,n$ are $\{0,3,4,6,6\}$. This violates the concavity because the change from $k=1$ to $k=2$ is only $1$, but the change from $k=2$ to $k=3$ is $2$. 
}
\label{fig:sets}
\end{figure}

Interval stabbing is a special case of the more general hitting set problem \cite{vinterbo2000hitting}, where the subsets to be hit are not restricted to contiguous intervals. In general, the corresponding cardinality-constrained hitting set problem does not have the same concavity property (see Figure~\ref{fig:sets}). This suggests that the concavity arises from the particular structure of intervals.

The term interval stabbing is also used in computational geometry to describe a data structure problem, where we need to organize a set of intervals to handle stabbing queries efficiently \cite{schmidt2009query}. The computational complexity of the stabbing or hitting set problems depends on the structure of points and subsets that we need to hit. More broadly, stabbing and hitting problems have been studied for various geometric objects, including two-dimensional rectangles \cite{kovaleva2006rectangle} and disks in the plane \cite{mustafa2010improved,bus2018geometric}. Such generalizations are computationally difficult and are often studied from the perspective of approximation algorithms. Other variants impose additional restrictions on the points that can be selected. For example, in partitioned interval stabbing, the candidate points are divided into subsets, and only entire subsets may be selected \cite{acharyya2021partitioned}. The integrality gaps of the hitting set and related problems have also been studied in the worst-case \cite{lovasz1975integrality} and random-input setting \cite{arpino2024random}. The related interval cover problem is a special case of the famous set cover problem. Its cardinality-constrained version is also known as the maximum coverage problem \cite{hochbaum1997approximating}. Extensions include multicover problems, in which objects have specified coverage requirements \cite{xu2024multicover}.

\section{Problem definition}

We denote by $[n]$ the set $\{1, 2, \ldots, n\}$,
and by $\IversonBracket{P}$ the value 1 if the condition $P$ is true, or 0 otherwise.

Consider the following optimization problem.  We are given a sequence
of intervals $[\ell_j, r_j]$ for $j \in [N]$, and a positive weight $w_j > 0$
is assigned to each interval $j$.  We wish to select at most $k$ real numbers
so as to maximize $\sum_j w_j$ over all the intervals $j$ that contain at least one
selected number. We also say that the selected number \textit{stabs} or \textit{hits} those intervals that contain it,
and hence call this the \textit{interval hitting} problem.

More formally, for a given selection of numbers, say $S \subseteq \RR$,
we define its value as $v(S) := \sum_j w_j \IversonBracket{S \cap  [\ell_j, r_j] \not= \emptyset}$.
We then define $f(k) := \max \{ v(S) : |S| \le k \}$,
i.e.\ the greatest sum achievable by selecting at most $k$ numbers;
and we are then interested in the value $f(k)$ for some desired value of $k$.

It is readily apparent that in solving this problem it is enough to limit oneself 
to selections where only the starting points of the intervals are used, 
i.e.\ where $S \subseteq \{\ell_1, \ldots, \ell_N\}$.  To see this, consider any
selection $S$ that contains some $x$ which is not the starting point of any interval.
(1)~If $x$ lies to the left of all intervals, it contributes nothing towards $v(S)$;
the selection $S$ will get no worse if $x$ is replaced by $\min_j \ell_j$.  
(2)~Otherwise, consider the rightmost $\ell_j$ that is less than $x$.  Then no interval
begins on $(\ell_j, x]$; hence any interval containing $x$ also contains $\ell_j$; 
hence the selection $S$ will get no worse if $x$ is replaced by $\ell_j$.

Therefore, in dealing with the interval hitting problem, we may begin by
sorting the values $\ell_j$ in increasing order and removing duplicates; 
let the resulting sequence be $\lambda_1 < \lambda_2 < \cdots < \lambda_n$ for some $n \le N$.
We can then replace the interval $[\ell_j, r_j]$ by $[L_j, R_j]$ (still with the same weight $w_j$), 
where $L_j$ is the unique index for which $\lambda_{L_j} = \ell_j$
and $R_j$ is the maximum index $t$ for which $\lambda_t \le r_j$.  
We have thus obtained an integer version of the interval hitting problem
in which the endpoints of the intervals are integers from $[n]$ and
a selection of integers $S \subseteq [n]$ is equivalent to
the selection $\{ \lambda_i : i \in S \}$ in the original problem.  

Thus, from now on we will limit ourselves to the \textit{integer interval hitting} problem,
where our intervals $[\ell_j, r_j]$, $j \in [N]$, have integer endpoints $\ell_j, r_j \in [n]$
and we seek a selection $S$ of at most $k$ (for some $k \le n$) integers from $[n]$ that maximizes $v(S)$.
However, the interval weights $w_j > 0$ may still be non-integers.
%Without loss of generality, we will assume that the intervals $[\ell_i, r_i]$ have
%been numbered in increasing order of $\ell_i$, so that $1 \le \ell_1 \le \ell_2 \le \cdots \le \ell_N \le n$.

%

It is easily seen that calculating $f(k)$ can be framed as an integer linear programming problem.
For each $i \in [n]$, we introduce a binary variable $x_i$ to indicate whether the integer~$i$
has been selected.  For each $j \in [N]$, we define $s_j := \sum_i x_i \IversonBracket{\ell_j \le i \le r_j}$
as the number of selected integers that hit the interval $j$, and we then introduce a binary variable
$y_j$ to indicate whether the interval $j$ has been hit at least once or not.
Morever, since all $w_j$ are positive, the optimal solution does not change if, instead
of selecting \textit{at most} $k$ integers, we are required to select \textit{exactly} $k$
integers, and we will
adopt this latter variant because it will be more convenient for the proofs in the next section.
This leads to the following problem:
\begin{equation}
\begin{array}{rl}
\text{maximize~~} & \textstyle J(\bx, \by) := \sum_{j=1}^N w_j y_j \\
\text{subject to~~} & 0 \le x_i \le 1 \textrm{~for each~}i \in [n] \\
                     & y_j \le 1 \textrm{~and~} y_j \le s_j \textrm{~for each~}j \in [N] \\
                     & \textstyle \sum_{i=1}^n x_i = k.
\end{array}\label{eq:konkavnost-linProg}
\end{equation}
(Note that $s_j$ is not a variable in our linear programming problem, but merely
a convenient shorthand for a certain sum of $x_i$'s.  We could write it as a function,
$s_j(\bx)$, but we will drop the ``$(\bx)$'' for convenience.)
Because all $w_j$ are positive, every optimal solution to this problem will
have $y_j = \min\{1, s_j\}$ for all $j$; hence when referring to an optimal
solution, it is enough to specify $\bx$ alone, and even to think of $J$ as simply a function of $\bx$.

\newcommand{\LP}[1]{\textup{LP}({#1})}
\newcommand{\ILP}[1]{\textup{ILP}({#1})}

We will refer to (\ref{eq:konkavnost-linProg}) as $\ILP{k}$ when
conceived as an integer linear programming problem, and as $\LP{k}$
if we relax it to a plain linear programming problem in which the variables
$x_i$ and $y_j$ may have non-integer values.

\section{Integrality of the LP relaxation}

We will show that the LP relaxation admits an optimal solution in which all variables $x_i$ are integral. The proof is based on a perturbation argument: starting from a suitable optimal solution $\bx$, we consider changes of the form $\bx\,\pm\,\eps \bc$. Lemma~\ref{lemma:c-exists} establishes the existence of a nonzero direction $\bc$ that preserves feasibility under sufficiently small perturbations. In particular, the condition $\benat \bc = 0$ preserves the cardinality constraint, while the additional conditions ensure that the selection levels $s_j$ remain unchanged for all intervals with $s_j=1$. We use these properties in Lemma~\ref{lemma:integer-solution} to establish the existence of an optimal solution with binary selection variables.

\begin{lemma}
\label{lemma:c-exists}
Let $(\bx, \by)$, where $\bx \in (0, 1)^n$\kern-0.25em, be a feasible solution of $\LP{k}$ in which all $x_i$ are non-integers.
Then there exists a vector $\bc \in \RR^n - \{\bn\}$ such that $\benat \bc = 0$ and
that for each $j \in [N]$, if $s_j = 1$, then $\sum_i c_i \IversonBracket{\ell_j \le i \le r_j} = 0$.
\textup{(}For convenience, we will refer to the left-hand side of this last equation as $\delta_j$.
Note that each $\delta_j$ is the sum of several $c_i$\kern-0.2em's exactly analogously to
how the corresponding $s_j$ is the sum of the corresponding $x_i$\kern-0.2em's.\textup{)}
\end{lemma}

\begin{proof}
Let $m$ be the number of $j$'s for which $s_j = 1$.  Then the conditions of our lemma
constitute a homogeneous system of $m + 1$ linear equations with $n$ unknowns, $c_1, \ldots, c_n$,
and the lemma asserts that this system has a nontrivial solution $\bc \not= \bn$.

For each $j$ where $s_j = 1$, our system has the equation $\delta_j = 0$.  
If two of these equations are identical, i.e.\ if the exact same terms $c_i$ occur on their left-hand
side, we can delete one of them without affecting the solutions of the system.

Can there be two equations, $\delta_j = 0$ and $\delta_{j'} = 0$, where the set of terms $c_i$
occurring in $\delta_j$ is a strict subset of that occurring in $\delta_{j'}$?  This means
that the set of terms $x_i$ occurring in $s_j$ is also a strict subset  % QW "strict" inserted
of that occurring in $s_{j'}$
(note that the sums $\delta_j$ and $\delta_{j'}$ are exactly analogous to $s_j$ and $s_{j'}$, except
that they use $c_i$ instead of $x_i$ as their terms).  But this means that $s_j \not= s_{j'}$
(because all $x_i$ are strictly positive), so they cannot both equal 1, so they cannot both
contribute an equation to our system.  

\iffalse
###
#
 #
  #
intervali: 1..3, 1..1, 2..2, 3..3       N = 4
števila: n = 3 
   s_1 = x_1 + x_2
   s_2 = x_1 + x_3
   s_3 = x_1 + x_4
\fi

Hence when we compare any two equations $\delta_j = 0$ and $\delta_{j'} = 0$ of our system,
each of them has some term $c_i$ that is not present in the other.  
Because of this, if two intervals $j$ and $j'$ have the same left endpoint, $\ell_j = \ell_{j'}$, then
they cannot both contribute an equation to our system.
Moreover, if two intervals $j$ and $j'$, where $\ell_j < \ell_{j'}$, both
contribute an equation to our system, it follows that $r_j < r_{j'}$,
otherwise the set of terms in $\delta_{j'} = 0$ would be a strict subset of the set of
terms in $\delta_j = 0$.  

Thus we can sort our equations $\delta_j = 0$ in (strictly) increasing order of
the corresponding $\ell_j$, and by doing so they will also be sorted in (strictly)
increasing order of the corresponding $r_j$.  We will denote by $\ell_{(t)}$ and $r_{(t)}$
the values $\ell_j$ and $r_j$ of the $t$'th equation in this order, for $t \in [m]$.

Note also that $\delta_j$ in each of these equations consists of
at least two terms (if it consisted of a single term, say $\delta_j = c_i$, then the corresponding
$s_j$ would also consist of a single term, $s_j = x_i$, hence $s_j$ would be $< 1$ and would not
contribute an equation to our system).  

Since each equation has at least two terms, $r_{(t)} > \ell_{(t)}$ holds for each $t \in [m]$.
Thus we have 
$1 \le \ell_{(1)} < r_{(1)} < r_{(2)} < \cdots < r_{(m)} \le n$.
Following these inequalities from left to right shows us that $t + 1 \le r_{(t)}$ for each $t \in [m]$,
% r(m) <= n
% r(m-1) <= n-1 
% ...
% r(m-u) <= n-u      pišimo t = m - u, torej u = m - t
% r(t) <= n - m + t
and going similarly from right to left shows that $r_{(t)} \le n - m + t$.
For $t = m$ this means that $m + 1 \le r_{(m)} \le n$, hence $m + 1 \le n$.

If $m + 1 < n$, there are more unknowns than equations, so the system certainly has nontrivial
solutions.  It remains to consider the case when $m + 1 = n$.  
We have seen that $t + 1 \le r_{(t)} \le n - m + t$, which now together with $m + 1 = n$ yields
$r_{(t)} = t + 1$.  Furthermore, from the fact that $1 \le \ell_{(1)} < \ell_{(2)} < \cdots < \ell_{(m)}$
it follows that $t \le \ell_{(t)}$ for each $t \in [m]$, which combined with $\ell_{(t)} < r_{(t)} = t + 1$
yields $\ell_{(t)} = t$.
Hence our system of $m + 1 = n$ equations can only have the following form:
\begin{gather*}
c_1 + c_2 = 0, \quad c_2 + c_3 = 0, \quad \ldots, \quad c_{n-1} + c_n = 0 \\
\text{and of course~} c_1 + c_2 + \ldots + c_n = 0.
\end{gather*}
If $n$ is even, this system has nontrivial solutions: choose any nonzero $c_1$
and then let $c_3, c_5, \ldots, c_{n-1}$ equal $c_1$, 
and let $c_2, c_4, \ldots, c_n$ equal $-c_1$.

This leaves the case when $n$ is odd.  The presence of the equation $c_i + c_{i+1} = 0$ in our system
means that some $s_j$ is of the form $x_i + x_{i+1}$ and has the value $1$.  
Moreover, since $\bx$ is a feasible solution of $\LP{k}$, it must satisfy the condition 
$\sum_{i=1}^n x_i = k$.  % QW popravljeno iz ^N na ^n
This last equation, since we know that $x_1 + x_2 = 1$, $x_3 + x_4 = 1$, etc.,
simplifies into $(n-1)/2 + x_n = k$; since $k$ is an integer and $n$ is odd, it follows that
$x_n = k - (n-1)/2$ must also be an integer, which contradicts the assumption of our lemma that 
$x_n$ is strictly between 0 and 1.  Hence the case of $m + 1 = n$ cannot occur when $n$ is odd.
\end{proof}

\begin{lemma}
\label{lemma:integer-solution}
$\LP{k}$ has an optimal solution in which all $x_i$ are integers.
\end{lemma}

\begin{proof}
The lemma trivially holds if $k = 0$ (when the only feasible solution is $\bx = \bn$)
or if $k = n$ (when the only feasible solution is $\bx = \bena$).  For cases when
$0 < k < n$, we will prove the lemma by contradiction.  
Suppose that there exists an instance of $\LP{k}$
in which every optimal solution has some non-integral $x_i$.  If there are several such
instances, take one with the least $k + n + N$ (we will subsequently refer to this sum
as the \textit{rank} of the instance).

\iffalse % True but perhaps not necessary?
			Consider, in our chosen instance of $\LP{k}$, two adjacent values $a_j$ and $a_{j+1}$.
			If they were both covered by the exact same set of intervals $[\ell_i, r_i]$, then
			the corresponding $s_j$ and $s_{j+1}$ would be identical, which means that in any optimal
			solution $y_j$ and $y_{j+1}$ would also be equal.  If we now constructed a new instance
			of $\LP{k}$ by merging $a_j$ and $a_{j+1}$ into a single value $a_j + a_{j+1}$ and
			adjust the boundaries of the intervals $[\ell_i, r_i]$ appropriately, the new instance
			would have the same set of optimal solutions $\bx$ as the original instance, with
			the same value of the cost function, but its $A + N$ would be lower (by~1) than that of the
			original instance, contracticting the assumption that our original instance was the
			one with the least $A + N$ amongst the instances for which all optimal solutions have
			some non-integral $x_i$.  
\fi

Consider an optimal solution $\bx$ of our instance of $\LP{k}$.  
We will begin by demonstrating that if any element of $\bx$ were an integer,
then a new instance could be constructed in which every optimal solution likewise
had some non-integral $x_i$, but whose rank would be lower than that of our original
instance, which is impossible considering how we chose the original instance to begin with.

(1)~Suppose that some $x_t$ had the value 0.  Then we could construct a new instance of $\LP{k}$
by changing each interval $[\ell_j, d_j]$ of our current instance into a new interval
$[\ell'_j, r'_j]$, defined as follows: $\ell'_j := \ell_j - \IversonBracket{\ell_j > t}$
and $r'_j := r_j - \IversonBracket{r_j \ge t}$; the weights of the intervals should remain unchanged.  
In the old instance, the set of candidates from which $k$ integers had to be selected was $[n]$; 
in the new instance, it is as if $t$ was deleted from the set of candidates, 
the candidates from $t+1$ to $n$ were decreased by 1, and the boundaries
of the intervals were suitably adjusted.  If any interval in the old instance was of the form $[t, t]$,
it should be deleted altogether from the new instance.  We will denote the cost function of the
new instance by $\hat{J}$, reserving $J$ for the cost function of the original instance.  
Now $\hat{\bx} := (x_1, \ldots, x_{t-1}, x_{t+1}, \ldots, x_n)$
is a feasible solution of the new instance, with the same value of the cost function as $\bx$ in the
old instance: $\hat{J}(\hat{\bx}) = J(\bx)$. 
Note that the new instance has a lower rank than the old one (because there are only $n-1$ candidates, the number of intervals is
$\le N$ and the $k$ is the same as before);
hence, by the way we have chosen our old instance, we can assume that the new
instance has some optimal solution $\hat{\bx}^\star$ where all its elements $\hat{x}^\star_i$ are integers;
if we augmented $\hat{\bx}^\star$ by inserting $x_t = 0$ at the appropriate index, 
the resulting augmented solution, say $\bx_a := (\hat{x}^\star_1, \ldots, \hat{x}^\star_{t-1}, 0,
\hat{x}^\star_{t+1}, \ldots, \hat{x}^\star_n)$, in which all $x_i$ are integers, 
would be a feasible solution of the old instance, again with $J(\bx_a) = \hat{J}(\hat{\bx}^\star)$. 
Combined with the fact that $\bx$ was an optimal solution of the old instance, we have
$J(\bx) \ge J(\bx_a) = \hat{J}(\hat{\bx}^\star) \ge \hat{J}(\hat{\bx}) = J(\bx)$,
hence $\bx_a$ is also an optimal solution of the old instance, which contradicts
the assumption that the old instance had no integral optimal solutions.  
% Thus we may henceforth assume that in $\bx$, no element $x_i$ equals 0.

\iffalse
			% Mislim, da tole tudi za prejšnjo verzijo dokaza, kjer smo govorili o
			% problemu pokrivanja števil z izbiranjem k intervalov, ni v redu, ker ne
			% dopusti možnosti, da x_{i'} prej doseže 1 (preseči pa je ne sme), preden x_i pade na 0.
			% Treba bi se bilo vrniti k starejši verziji iz kakšnega od prejšnjih mailov.
			% V vsakem primeru pa to za naš novi problem (prebadanje intervalov) ni relevantno.
			(2)~Let $H_i := \{ j \in [N] : \ell_j \le i \le r_j \}$ be the set of intervals containing $i$.
			Suppose that $H_i \subseteq H_{i'}$ were true for some $i \not= i'$.  
			We have just seen that no element of our optimal solution equals 0, hence $x_i$ and $x_{i'}$ are both nonzero.
			We can decrease $x_i$ and increase $x_{i'}$ by the same amount, until 
			$x_i$ drops to 0; the solution remains feasible and 
			its cost function doesn't decrease (because each $a_j$ that was covered by interval $i$
			is also covered by interval $i'$, hence if $x_i$ appears in any $s_j$, it appears 
			as a part of $x_i + x_{i'}$, hence no $s_j$ decreases and hence the cost function
			doesn't decrease either); hence the new solution is still optimal.  
			But since $x_i$ has dropped to 0, we are in contradiction with paragraph~(1);
			hence it is impossible for an interval $i$ to be a subset of some other interval $i'$.
\fi

(2)~Suppose that some $x_t$ had the value 1.  We can construct a new instance of $\LP{k-1}$
by deleting all the intervals containing $t$ and then proceeding as in (1) to remove $t$
from the set of candidates and suitably adjust the interval boundaries; moreover, the new
instance will now require us to select only $k-1$ numbers instead of $k$ numbers.  
Again $\hat{\bx} := (x_1, \ldots, x_{t-1}, x_{t+1}, \ldots, x_n)$ is a feasible
solution of the new instance, with the cost function $\hat{J}(\hat{\bx}) = J(\bx) - \Delta$
for $\Delta := \sum_j w_j \IversonBracket{\ell_j \le t \le r_j}$, this being the
total weight of the intervals which were hit by $t$ in the old instance
but which are absent from the new instance.  The new instance again has a lower rank
than the old instance, hence it must have some optimal 
solution $\hat{\bx}^\star$ where all its elements $\hat{x}^\star_i$ are integers;
if we augmented $\hat{\bx}^\star$ it by inserting $x_t = 1$ at the appropriate index, 
the resulting augmented solution, say $\bx_a := (\hat{x}^\star_1, \ldots, \hat{x}^\star_{t-1}, 1,
\hat{x}^\star_{t+1}, \ldots, \hat{x}^\star_n)$, in which all $x_i$ are integers, 
would be a feasible solution of the old instance, with $J(\bx_a) = \hat{J}(\hat{\bx}^\star) + \Delta$. 
Combined with the fact that $\bx$ was an optimal solution of the old instance, we have
$J(\bx) \ge J(\bx_a) = \hat{J}(\hat{\bx}^\star) + \Delta \ge \hat{J}(\hat{\bx}) + \Delta = J(\bx)$,
hence $\bx_a$ is also an optimal solution of the old instance, which contradicts
the assumption that the old instance had no integral optimal solutions.  
%Thus we may henceforth assume that in $\bx$, no element $x_i$ equals 1.

In view of properties (1) and (2), we may now assume that every $x_i$ lies
strictly between 0 and 1.  By Lemma~\ref{lemma:c-exists}, there exists a $\bc \not= \bn$
such that $\benat \bc = 0$ and that $\delta_j = 0$ for each $j$ where $s_j = 1$
(with $\delta_j$ being defined as $\sum_i c_i \IversonBracket{\ell_j \le i \le r_j}$).

Suppose we took a small $\eps > 0$ and changed our solution from $\bx$ to 
$\hat{\bx} := \bx + \eps \bc$.  Since all $x_i$ were from $(0, 1)$,
the new $\hat{x}_i$ will be in $[0, 1]$ if $\eps$ is small enough;
moreover because of $\benat \bc = 0$, we will have $\benat \hat{\bx} = \benat \bx = k$;
hence the new solution is still feasible.  How does the cost function change
because of this move?  We will denote by $s_j$ the values before the change and
by $\hat{s}_j := s_j + \eps \delta_j$ those after the change; likewise we have
$y_j = \min\{1, s_j\}$ before the change and $\hat{y}_j = \min\{1, \hat{s}_j\}$
after the change.  Thus the change from $s_j$ to $\hat{s}_j$ causes the cost function
to change by $(\hat{y}_j - y_j) w_j$.
(\textit{a})~For those $j$ where $s_j < 1$,
if we choose a sufficiently small $\eps > 0$, the new $\hat{s}_j$ will be $\le 1$,
hence $\hat{y}_j = \hat{s}_j = s_j + \eps \delta_j = y_j + \eps \delta_j$,
hence the cost function will change by $w_j \eps \delta_j$.
(\textit{b})~For those $j$ where $s_j = 1$, we know that $\delta_j = 0$,
hence $\hat{y}_j = \hat{s}_j = s_j = y_j = 1$ and the cost function will not change.
(\textit{c})~For those $j$ where $s_j > 1$, if we choose a sufficiently small $\eps > 0$,
the new $\hat{s}_j$ will be $\ge 1$, hence
$\hat{y}_j = 1 = y_j$ and the cost function will not change.

If, instead of moving from $\bx$ to $\bx + \eps \bc$, we moved to $\bx - \eps \bc$,
we would find that in case (\textit{a}) the cost function would change
by $- w_j\eps \delta_j$ while in cases (\textit{b}) and (\textit{c}) it would remain unchanged.

Thus, if we sum up the above observations over all $j \in [N]$,
we see that the total effect of these two moves is that in one of them the cost function
increases by some amount and in the other it decreases by the same amount.
However, the cost function cannot actually increase because $\bx$ was already
an optimal solution; hence it follows that the amount by which the moves from $\bx$
to $\bx \pm \eps \bc$ increase or decrease the cost function can only be 0.
Hence whichever of the two moves we choose, the new solution after the move 
will remain optimal.

We can slowly increase $\eps$ (the length of the move) until one of the
following happens: either some $x_i$ drops to 0, or some $x_i$ increases to 1,
or some $s_j$ that was initially different from 1 now reaches the value of 1.
But the first two of these three outcomes are impossible, as we have seen
in paragraphs (1) and (2) above that no optimal solution of our instance of $\LP{k}$
has any integer $x_i$.  Hence the effect of our move, for a suitable choice of $\eps$,
is that one $s_j$ will have moved from a value $s_j \not= 1$ to the value $s_j = 1$, whereas those $s_j$'s which
were already equal to 1 will have remained unchanged.

We can now make a new move, with a new choice of $\bc$ and $\eps > 0$,
and cause yet another $s_j$ to reach the value~1.  By repeating this, we 
eventually obtain a solution $\bx$ 
(still optimal and still with all $x_i$ strictly between 0 and 1)  
in which \textit{all} $s_j$ are equal to~1.  But Lemma~\ref{lemma:c-exists}
still applies; we can choose another nonzero $\bc$ and $\eps > 0$
for which $\bx + \eps \bc$ remains a feasible and optimal solution.
But now, since moving in the direction $\bc$ does not change \textit{any}
$s_j$, we can certainly choose an $\eps$ large enough that some $x_i$
will reach the value 0 or 1.  Hence there exists an optimal solution
with some integer $x_i$, but this, as we have seen in paragraphs (1) and~(2),
is in contradiction with the way we chose our instance of $\LP{k}$ to begin with.
\end{proof}

\begin{corollary}
\label{corollary:lp-phi}
The value of the cost function in the optimal solution of $\LP{k}$ is $f(k)$.
\end{corollary}

\begin{proof}
Lemma~\ref{lemma:integer-solution} has shown that $\LP{k}$ has an optimal solution $\bx$
in which all $x_i$ are integers; hence this is also a feasible solution of
$\ILP{k}$, with the same value of the cost function.  This value, $J(\bx)$, must therefore be $\le f(k)$,
since, by definition, $f(k)$ is the optimal value of the cost function in
$\ILP{k}$.  But since every feasible solution of $\ILP{k}$ is also a feasible
solution of $\LP{k}$, the optimal value of $\LP{k}$ (i.e.\ $J(\bx)$) 
cannot be less than that of $\ILP{k}$ (i.e.\ $f(k)$); hence it follows that $J(\bx) = f(k)$.
\end{proof}

\section{Concavity}

We have shown that the LP relaxation admits an optimal solution with integral selection variables. We now use this result to establish the concavity of the optimal value function.
Exploiting integral optimal solutions of LP relaxations to establish properties of the corresponding combinatorial optimization problems is a standard approach \cite{schrijver1986relax}.
First, we show that the optimal value $J(\bx)$ of the relaxed problem $\LP{k}$ is concave as a function of $k$. The integrality result then allows us to conclude that the optimal value function $f(k)$ of the integer problem $\ILP{k}$ is concave as well.

\begin{theorem}
\label{theorem:f-concave} % was: lemma:phi-concave
For any given sequence of positive numbers $w_1, \ldots, w_N$ and list of intervals $[\ell_j, r_j]$, $j \in [N]$,
the function $f(k)$ is concave.
\end{theorem}

\begin{proof}
Let $k_1, k_2, k_3$ be three integers such that $k_1 < k_2 < k_3$, and let $\mu := (k_2 - k_1) / (k_3 - k_1)$.
To prove concavity, we must show that $f(k_2) \ge (1-\mu) f(k_1) + \mu f(k_3)$.

For our particular list of intervals $[\ell_j, r_j]$ and their weights $w_j$ (for all $j \in [N]$),
the corresponding problems $\LP{k}$ differ only in the value of $k$ used in the
constraint $\benat \bx = k$ (and hence in the set of feasible solutions).
Notably, they all share the same cost function $J(\bx)$.
For $t = 1, 2, 3$, let $\bx^{(t)}$ be any optimal solution of $\LP{k_t}$;
and let $\bs^{(t)}$ and $\by^{(t)}$ be the vectors of the values $s_j$ and $y_j$ corresponding
to $\bx^{(t)}$.

Define $\hat{\bx} := (1-\mu) \bx^{(1)} + \mu \bx^{(3)}$.
Note that $\mu \in (0, 1)$; hence, since $x^{(1)}_i$ and $x^{(3)}_i$ are from $[0, 1]$,
it follows that $\hat{x}_i = (1-\mu) x^{(1)}_i + \mu x^{(3)}_i$ is also from $[0, 1]$.
Moreover, $\benat \hat{\bx} =
\benat \big((1-\mu)\bx^{(1)} + \mu \bx^{(3)}\big)
= (1-\mu) \benat \bx^{(1)} + \mu \benat \bx^{(3)}
= (1-\mu) k_1 + \mu k_3 = k_2$.  Hence, $\hat{\bx}$ is a feasible solution of $\LP{k_2}$.

From the way we have defined $\hat{\bx}$, it follows that the corresponding $\hat{s}_j$ form
the vector $\hat{\bs} = (1 - \mu) \bs^{(1)} + \mu \bs^{(3)}$.  As for the corresponding $\hat{y}_j$,
note that $\Upsilon(u) := \min\{1, u\}$ is a concave function;
hence, $\hat{y}_j = \Upsilon(\hat{s}_j) = \Upsilon\big( (1-\mu) s^{(1)}_j + \mu s^{(3)}_j \big)
\ge (1-\mu) \Upsilon(s^{(1)}_j) + \mu \Upsilon(s^{(3)}_j) = (1 - \mu) y^{(1)}_j + \mu y^{(3)}_j$.
And hence, $J(\hat{\bx}) = \sum_j w_j \hat{y}_j \ge (1-\mu) \sum_j w_j y^{(1)}_j + \mu \sum_j w_j y^{(3)}_j
= (1-\mu) J(\bx^{(1)}) + J(\bx^{(3)})$.  

Since $\bx^{(t)}$ is an optimal solution of $\LP{k_t}$, it follows from
Corollary~\ref{corollary:lp-phi} that $J(\bx^{(t)}) = f(k_t)$.  
The solution $\hat{\bx}$ of $\LP{k_2}$ cannot, of course, 
be better than the optimal solution of that problem, which is $\bx^{(2)}$.
Thus we have
$f(k_2) = J(\bx^{(2)}) \ge J(\hat{\bx}) \ge (1-\mu) J(\bx^{(1)}) + J(\bx^{(3)})
= (1 - \mu) f(k_1) + \mu f(k_3)$,
which completes the proof.
\end{proof}

\iffalse
	\begin{theorem}
	\label{theorem:f-concave}
	For any given sequence of positive numbers $w_1, \ldots, w_N$ and list of intervals $[\ell_j, r_j]$, $j \in [N]$,
	the function $f(k)$ is concave.
	\end{theorem}

	\begin{proof}
	By Lemma~\ref{lemma:phi-concave}, $\phi(k)$ is concave; hence, as $k$ grows, the slope
	of $\phi$ never increases, but only decreases or remains constant.  For as long as 
	the slope of $\phi$ is positive, the values $\phi(k)$ increase and each new value is
	the largest so far; hence $f(k) = \max_{\kappa \le k} \phi(\kappa) = \phi(k)$,
	so $f$ has the same slope as $\phi$, so the slope of $f$ decreases just like that of $\phi$ does.  
	But at the point when the slope of $\phi$ drops
	below 0, $\phi(k)$ begins to decrease and thereafter never stops decreasing,
	so $f(k)$ will remain constant at the highest value that $\phi$ achieved before it
	began to decrease.  Thus, the last change in the slope of $f$ will be from some
	nonnegative slope (the last slope of $\phi$ before it began to decrease) to 0.
	Thus, the slope of $f$ never increases, which means that $f$ is concave.
	\end{proof}
\fi

\section{Interval cover}

We will now consider another closely related optimization problem, the \textit{interval cover}
problem.  We are given a sequence of values $w_1, \ldots, w_N$ and a sequence of
intervals $[\ell_i, r_i]$, $i \in [n]$, where $\ell_i$ and $r_i$ are integers from
the range $1 \le \ell_i \le r_i \le N$.  The goal is to select up to $k$ intervals
so as to maximize the sum of covered values $\sum_j w_j$ , where $j$ goes over the union of the selected intervals.

More formally, for a given selection of intervals, say $S \subseteq [n]$, we
define its value as $\hat{v}(S) := \sum_j w_j \IversonBracket{\exists i \in S : \ell_i \le j \le r_i}$.
We then define $\hat{f}(k) := \max\{ \hat{v}(S) : |S| \le k\}$, very much like
for the interval hitting problem.  

To demonstrate that $\hat{f}(k)$ is concave, we will reduce the interval cover
problem to the interval hitting problem.  Note that in the interval cover problem,
if $[\ell_i, r_i]$ is a subset of $[\ell_{i'}, r_{i'}]$ for some $i \not= i'$,
there is never any benefit to selecting $i$ (since we could select $i'$ instead)
and the optimal solution does not change if we delete interval $i$ from the problem.  
After this no two intervals begin at the same point (or one would be a
subset of the other); moreover, if one interval begins to the left of another,
it certainly also ends to the left of the other one (or this other interval would be its subset).  
Hence if we now order the intervals in increasing value of $\ell_i$,
the corresponding $r_i$ will also form an increasing sequence: $\ell_1 < \ell_2 < \cdots < \ell_n$
and $r_1 < r_2 < \cdots < r_n$.  

Take three intervals $i, i', i''$ such that $i < i' < i''$.  Let $j \in [N]$ be any number
that belongs to both $i$ and $i''$; since $j$ belongs to $i$, we have $j \le r_i < r_{i'}$,
and since it belongs to $i''$, we have $\ell_{i'} < \ell_{i''} \le j$; hence $j$ also belongs to $i'$.
Thus, for any $j \in [N]$, the set of intervals containing $j$ is itself something like an interval:
$j$ belongs to the intervals $[\ell_i, r_i]$ for all $i$ from the range $L_j \le i \le R_j$,
defined by $L_j := \min \{i : r_i \ge j\}$ and $R_j := \max \{i : \ell_i \le j \}$.  

We can now use the $N$ intervals $[L_j, R_j]$, with the corresponding weights $w_j$,
as an instance of the interval hitting problem.  Since the endpoints $L_j$ and $R_j$ are all
from $[n]$, any optimal solution of the new problem will also select only integers from $[n]$.
For any such selection $S \subseteq [n]$, its value (in the context of the interval hitting
problem) is 
\begin{align*}
v(S) &= \textstyle\sum_j w_j \IversonBracket{S \cap [L_j, R_j] \not= \emptyset} \\
&= \textstyle\sum_j w_j \IversonBracket{\exists i \in S : L_j \le i \le R_j} \\
&= \textstyle\sum_j w_j \IversonBracket{\exists i \in S : j \in [\ell_i, r_i]} \\
&= \textstyle\sum_j w_j \IversonBracket{\exists i \in S : \ell_i \le j \le r_i} \\
&= \hat{v}(S),
\end{align*}
i.e.\ the same as its value in the context of the interval cover problem.
Thus any candidate selection of no more than $k$ intervals for our interval cover
problem is also a candidate selection of no more than $k$ integers for our interval hitting problem
(and vice versa), with the same value of the cost function; hence the optimal solution of both problems
is the same, $\hat{f}(k) = f(k)$.  By Theorem~\ref{theorem:f-concave}, $f$ is concave,
therefore $\hat{f}$ is also concave.

\section{An alternative proof of integrality}

As an alternative to the more direct approach of
lemmas \ref{lemma:c-exists} and~\ref{lemma:integer-solution}, 
the existence of integer optimal solutions of $\LP{k}$ 
can also be proven by relying on certain known properties of totally 
unimodular matrices.  A matrix $M$ is defined as \textit{totally unimodular} 
iff each square submatrix of $M$ has a determinant from $\{0, 1, -1\}$.

\iffalse
	We have already observed that, since all the weights $w_j$ in our definition 
	of the interval hitting problem are greater than 0, there is no difference, 
	as far as the optimal solution is concerned, between having to select 
	\textit{exactly} $k$ numbers and having to select \textit{at most} $k$ numbers.  
	Thus the optimal solutions of our $\LP{k}$ do not change if we replace
	the condition $\sum_{i=1}^n x_i = k$ with $\sum_{i=1}^n x_i \le k$.
	After this change, we can rewrite the problem in matrix form thus:
\fi
We can rewrite the problem $\LP{k}$ in matrix form thus:
$$\text{maximize~} J(\bz) := \hat{\bw}^\top \bz \text{~subject to~} A\bz \le \bb,$$
where
$$
A := \begin{bmatrix}
-I_n & 0_{n \times N} \\
 I_n & 0_{n \times N} \\
 0_{N \times n} & I_N \\
-S & I_N \\
\bena^{\top}_n & \bn^{\top}_N \\
-\bena^{\top}_n & \bn^{\top}_N
\end{bmatrix},
\bb := \begin{bmatrix} 
 \bn_n \\
 \bena_n \\
 \bena_N \\
 \bn_N \\
 k \\
 -k 
\end{bmatrix},
\bz := \begin{bmatrix}
 \bx \\
 \by
\end{bmatrix},
\text{~and~}
\hat{\bw} := \begin{bmatrix}
 \bn_n \\
 \bw
\end{bmatrix}
$$
and where $S$ is a matrix of $N$ rows and $n$ columns with the elements
$S_{ji} := \IversonBracket{\ell_j \le i \le r_j}$ (so that our sums $s_j$ 
constitute the vector $\bs = S \bx$).  

A matrix $M$ is an \textit{interval matrix} iff it has at most one contiguous block
of 1's in each column, and the rest of the entries are 0's.  Every interval matrix is
totally unimodular \cite[p.~279]{schrijver1986theory}.

Since $S$ has a single contiguous block of 1's in each row, and 0's elsewhere,
it follows that $S^\top$ is an interval matrix; and if we add two columns of 1's,
the resulting $[S^\top, \bena_n, \bena_n]$ is also an interval matrix and therefore
totally unimodular.

From the definition of total unimodularity and from the properties of determinants it
readily follows
that the following operations preserve the total unimodularity of a matrix:
multiplying a row by $-1$; transposing the matrix; inserting 
a new row or column in which at most one element is $\pm 1$ and the others are $0$.
By the means of such operations we can easily change $[S^\top, \bena_n, \bena_n]$ into $A$,
therefore $A$ is also totally unimodular.

% We now have the problem
%    maximize J(z) := w^T z       (we will write w instead of \hat{w} for convenience)
%    such that Az <= b
% From Schrijver's Theorem 19.1 (p. 266) we know that the polyhedron
%   P := { w : Aw <= b }
% is integral.  
%
% We want to prove that the optimal solution of our problem remains unchanged
% if we restricts ourselves to z's in which all elements are integers.
% In principle that is what Corollary 19.1a in Schrijver (p. 267) is for,
% but that corollary requires the weights in w to also be integers.
% I don't understand the theory behind it enough to see why this is necessary,
% but it seems plain enough that 

% Suppose this were not the case.  Let z \in P be a feasible solution with
% some non-integer components and whose w^T z is greater than that of any
% feasible solution with purely integer components.  But since P is integral,
% any feasible solution z \in P can be expressed as a convex combination 
% of some integer feasible solutions from P, say z_1, ..., z_t, such that    
%    z = sum_i lambda_i z_i   for some real values lambda_i >= 0 where sum_i lambda_i = 1.
% Among the vector z_1, ..., z_t, let z_j be the one with the greatest value J(z_j).
% Then J(z_j) = w^T z_j = 1 * (w^T z_j) = (sum_i lambda_i) (w^T z_j)
% = sum_i (lambda_i w^T z_j) 
% >= sum_i (lambda_i w^T z_i)      % because for each i, we have w^T z_j >= w^T z_i -- that's how we chose j
% = sum_i w^T (lambda_i z_i)
% = w^T (sum_i lambda_i z_i) 
% = w^T z = J(z).
% 
% Thus our non-integer solution z cannot actually be strictly better than 
% the integer solution z_j.

Since our linear programming problem has the constraints $A \bz \le \bb$
with a totally uniform $A$ and an integer vector $\bb$, 
we can be sure that it has some integral optimal solution 
\cite[p.~267]{schrijver1986theory}.

\section{Conclusion}

We have shown that the optimal value function $f(k)$ of the cardinality-con\-strain\-ed interval stabbing problem is concave in the cardinality constraint $k$. The proof establishes that the LP relaxation admits an optimal solution with integral selection variables, and uses this result to show that the optimal value function is concave. We further show that the same property holds for the cardinality-constrained interval cover problem by reducing it to the interval stabbing problem.

These concavity results provide a basis for applying Lagrangian optimization to these constrained problems \cite{jain2001lagrange}. By introducing a penalty $\lambda$ for each selected element, we obtain a related unconstrained problem in which the objective is the total weight of covered intervals minus $\lambda$-times the number of selected elements. Larger penalties lead to solutions with fewer selected elements, and smaller penalties lead to solutions with more selected elements. The concavity of $f(k)$ ensures that suitable penalty values support the optimal value function at a given cardinality constraint. It allows us to binary search for a penalty whose unconstrained problem yields a solution that matches the cardinality constraint. This approach can be used to solve the constrained problem with an additional logarithmic factor in the running time.

%\begin{acknowledgements}
%\end{acknowledgements}


\bibliographystyle{spmpsci}      % mathematics and physical sciences
\bibliography{intervals.bib}   % name your BibTeX data base

@article{jain2001lagrange,
  title={Approximation algorithms for metric facility location and k-median problems using the primal-dual schema and Lagrangian relaxation},
  author={Jain, Kamal and Vazirani, Vijay V},
  journal={Journal of the ACM (JACM)},
  volume={48},
  number={2},
  pages={274--296},
  year={2001},
  publisher={ACM, New York, NY, USA}
}

@article{mustafa2010improved,
  title={Improved results on geometric hitting set problems},
  author={Mustafa, Nabil H and Ray, Saurabh},
  journal={Discrete \& Computational Geometry},
  volume={44},
  number={4},
  pages={883--895},
  year={2010},
  publisher={Springer}
}

@InProceedings{arpino2024random,
  author =	{Arpino, Gabriel and Dmitriev, Daniil and Grometto, Nicolo},
  title =	{{Greedy Heuristics and Linear Relaxations for the Random Hitting Set Problem}},
  booktitle =	{Approximation, Randomization, and Combinatorial Optimization. Algorithms and Techniques (APPROX/RANDOM 2024)},
  pages =	{30:1--30:22},
  series =	{Leibniz International Proceedings in Informatics (LIPIcs)},
  ISBN =	{978-3-95977-348-5},
  ISSN =	{1868-8969},
  year =	{2024},
  volume =	{317},
  editor =	{Kumar, Amit and Ron-Zewi, Noga},
  publisher =	{Schloss Dagstuhl -- Leibniz-Zentrum f{\"u}r Informatik},
  address =	{Dagstuhl, Germany}
}

@article{bus2018geometric,
  title={Practical and efficient algorithms for the geometric hitting set problem},
  author={Bus, Norbert and Mustafa, Nabil H and Ray, Saurabh},
  journal={Discrete Applied Mathematics},
  volume={240},
  pages={25--32},
  year={2018},
  publisher={Elsevier}
}

@InProceedings{schmidt2009query,
author="Schmidt, Jens M.",
editor="Dong, Yingfei
and Du, Ding-Zhu
and Ibarra, Oscar",
title="Interval Stabbing Problems in Small Integer Ranges",
booktitle="Algorithms and Computation",
year="2009",
publisher="Springer Berlin Heidelberg",
address="Berlin, Heidelberg",
pages="163--172",
isbn="978-3-642-10631-6"
}

@inbook{erickson2019algorithms,
author = {Erickson, Jeff},
year = {2019},
month = {06},
chapter = {4},
title = {Algorithms},
isbn = {978-1-792-64483-2},
}

@InProceedings{xu2024multicover,
author="Xu, Yuxuan
and Ran, Yingli
and Zhang, Zhao",
editor="Ghosh, Smita
and Zhang, Zhao",
title="Approximation Algorithm for the Maximum Interval Multi-cover Problem",
booktitle="Algorithmic Aspects in Information and Management",
year="2024",
publisher="Springer Nature Singapore",
address="Singapore",
pages="28--34",
isbn="978-981-97-7798-3"
}

@article{hochbaum1997approximating,
  title={Approximating covering and packing problems: set cover, vertex cover, independent set, and related problems},
  author={Hochbaum, Dorit S},
  journal={Approximation algorithms for NP-hard problems},
  pages={94--143},
  year={1997},
  publisher={PWS Publishing}
}

@article{kovaleva2006rectangle,
author = { Kovaleva, Sofia and  Spieksma, Frits C. R.},
title = {Approximation Algorithms for Rectangle Stabbing and Interval Stabbing Problems},
journal = {SIAM Journal on Discrete Mathematics},
volume = {20},
number = {3},
pages = {748-768},
year = {2006},
}

@article{vinterbo2000hitting,
title = {Minimal approximate hitting sets and rule templates},
journal = {International Journal of Approximate Reasoning},
volume = {25},
number = {2},
pages = {123-143},
year = {2000},
issn = {0888-613X},
author = {Staal Vinterbo and Aleksander Øhrn},
}

@InProceedings{acharyya2021partitioned,
author="Acharyya, Ankush
and Keikha, Vahideh
and Majumdar, Diptapriyo
and Pandit, Supantha",
editor="Chen, Chi-Yeh
and Hon, Wing-Kai
and Hung, Ling-Ju
and Lee, Chia-Wei",
title="Constrained Hitting Set Problem with Intervals",
booktitle="Computing and Combinatorics",
year="2021",
publisher="Springer International Publishing",
address="Cham",
pages="604--616",
}

@article{lovasz1975integrality,
title = {On the ratio of optimal integral and fractional covers},
journal = {Discrete Mathematics},
volume = {13},
number = {4},
pages = {383-390},
year = {1975},
issn = {0012-365X},
author = {L. Lovász},
}

@book{schrijver1986theory,
title = {Theory of Linear and Integer Programming},
author = {Alexander Schrijver},
publisher = {John Wiley \& Sons},
year = {1986},
series = {Wiley-Interscience Series in Discrete Mathematics}
}

@article{schrijver1986relax,
title = {Polyhedral proof methods in combinatorial optimization},
journal = {Discrete Applied Mathematics},
volume = {14},
number = {2},
pages = {111-133},
year = {1986},
issn = {0166-218X},
author = {A. Schrijver}
}

\end{document}